\documentclass[reqno]{article}
\usepackage{amsthm} 
\usepackage{amstext}
\usepackage{amsbsy}
\usepackage{epsf}
\usepackage{bbm}
\usepackage{mathrsfs}
\usepackage{amscd}
\usepackage{amsmath}
\usepackage{amsfonts}
\usepackage{hyperref}
\usepackage{setspace}
\usepackage{latexsym}
\usepackage{amssymb}
\usepackage{fancyhdr}
\usepackage{times}
\usepackage{textcomp}
\usepackage [T1] {fontenc}
\usepackage{graphicx}
\usepackage{epsfig}
\usepackage[graphics,floats,textmath,displaymath,delayed]{preview}
\usepackage{url}

\newtheorem{theorem}{Theorem}

\newtheorem{proposition}{Proposition}

\theoremstyle{definition}

\theoremstyle{remark}

\numberwithin{equation}{section}

\newcommand{\R}{\mathbb{R}}

\newcommand{\N}{\mathbb{N}}

\newcommand{\Z}{\mathbb{Z}}
\newcommand{\T}{\mathbb{T}}

\newcommand{\q}{\boldsymbol{q}}

\newcommand{\F}{\mathcal{F}}

\newcommand{\non}{\nonumber}

\newcommand{\hf}{{_1\over^2}}
\newcommand{\p}{\boldsymbol{p}}

\newcommand{\supp}{\mathop{\hbox{{\rm Supp}}}}
\def\qq{ \begin{eqnarray} }
\def\qqq{ \end{eqnarray} }

\doublespace

\begin{document}



\begin{center}

\normalsize{\textbf{KAM theorem and Renormalization Group}}

\vspace{1cm}

\footnotesize{

\textrm{E. De Simone, A. Kupiainen } \\

\vspace{0.2cm}

\textit{Department of Mathematics and Statistics, P.O.Box 68 (Gustaf Hällströmin katu 2{b}) \\
Helsinki, 00014 , Finland} 
}

\end{center}
\vspace{0.5cm}

\begin{center}
\footnotesize{ ABSTRACT 
\vspace{0.3cm}

\parbox[c]{300pt}{We give an elementary proof of the analytic KAM theorem
by reducing it to a Picard iteration of a PDE with quadratic
nonlinearity, the so called Polchinski renormalization group 
equation studied  in quantum field theory.}}
\end{center}
\vspace{0.4cm}

\normalsize

\section{Introduction}

An integrable Hamiltonian system is
given in the action-angle coordinates
$(I, \theta)\in \T^d\times \R^d$ by a Hamiltonian $H_0(I)$
depending only on the actions. It generates the flow
\begin{equation}
\label{int}
\dot I=0, \ \ \dot\theta=\omega(I)
\end{equation}
where $\omega=\partial_I H_0$. The orbit $\theta(t)=\omega(I(0))t$
is quasiperiodic and spans the torus $I=I(0)$
for almost all $I(0)$ in a neighbourhood of a point
$I$ where the Hessian $D^2H_0$ is nondegenerate.
The KAM theorem 
constructs invariant tori in such a neighbourhood
when $H_0$ is perturbed by  a  Hamiltonian $\lambda H_1(I,\theta)$
suitably small in the analytic (\cite{kol}, \cite{ar})
or $C^k$  \cite{mos} category. These solutions
are found by a Newton iteration that constructs
coordinates where the flow takes the simple form \eqref{int}.

An alternative approach in the analytic case is a perturbation
expansion of the solutions in powers of $\lambda$. The resulting
Lindstedt series which converges as a consequence of
the KAM theorem \cite{kol} was proven to do so
by directly bounding the coefficients much later
by Eliasson \cite{eli} who showed the existence of
huge cancellations in the terms in the series. 
Subsequently Eliasson's work was simplified and extended
by Gallavotti  \cite{g1,g2,g3}, by Chierchia and Falcolini \cite{gf}
and by Bonetto, Gentile and Mastropitero \cite{bgm}. 
In particular Gallavotti {\it et. al.} revealed the
analogy of Lindstedt series to perturbative expansions in 
quantum field theory (QFT) and used ideas, in particular
the renormalization group (RG),  that they had 
developed for the latter to the former. 

These papers
in turn were an inspiration to the study \cite{bgk}
which developed a new iterative scheme, inspired by the RG. 
In this approach
the small denominators that plague the Lindstedt series
are studied scale by scale by introducing renormalized
problems for the unresolved scales. In the iterative scheme
the renormalized problems become simpler 
as the scale of the denominators is reduced.
This fact was shown to be a consequence of symmetries 
in the problem which fact provides an explanation
for the cancellations that Eliasson uncovered in his proof.
The renormalization group iteration was subsequently
extended to PDE's \cite{bks}, to  the
homoclinic splitting problem \cite{mikko} and also to
the case of a $C^k$ perturbation \cite{emil} where the Lindstedt series
is not convergent.

In this paper we return to the analytic KAM theorem and
give a proof of it using a RG where the scale is
changed continously. The renormalized problems
are given as solutions of a simple evolution PDE
with quadratic nonlinearity. This equation is
well known in QFT where it goes under the name
Polchinski equation \cite{pol}.  Although conceptually
this proof is just an infinitesimal version of the
one given in \cite{bgk} we feel it is worth writing
down for two reasons.

First, the KAM theorem is reduced to a very simple
Picard iteration with elementary
estimates. 

Secondly we wish to bring into attention the Polchinski
equation in this context. It is well known that as the
size $\lambda$ of the perturbation is increased the torus with a given
frequency $\omega$ disappears. For certain "scale invariant"
  $\omega$  (e.g.  in $d=2$ for $\omega=(1,\gamma)$ with  
$\gamma$ a "noble" irrational) the solution at the critical
$\lambda$ is believed to have "universal" scaling
properties \cite{kada,shenker}. Such universality presumably
would be related to fixed points of the RG equation
(composed with a scaling). It would be very interesting
to see if the RG equation possesses such scaling solutions.
The simplicity of the equation means that this hope might
not be entirely unrealistic.

\section{The RG Equation}

We will give the proof of KAM theorem in the special case
where the perturbation $ H_1$ is a function of the angles
$\theta$ only.  This case possesses all the difficulties of the
general one (for the RG formalism in the general case see
\cite{bgk}). Thus, given a real analytic function
$v:\T^d\to \R$ where $\T$ is the circle $\R/2\pi\Z$,
we look for quasiperiodic solutions to the equation
\begin{equation}
\label{newton}
\ddot \theta=-\lambda\partial v(\theta) 
\end{equation}
where $\lambda$ is a perturbation parameter. The solutions are constructed
by by fixing a vector $\omega\in\R^d$ whose components are independent over
the integers and looking for a function $X:\T^d\to \R^d$ s.t.
$$\theta(t)=\omega t+X(\omega t)$$
solves \eqref{newton}. This follows, if $X$ solves the equation
\begin{equation}
\label{main0}-(\omega\cdot\partial )^2X= U(X)
\end{equation}
where
\begin{equation}
\label{main1}
U(\theta,X) = \lambda \partial v(\theta + X(\theta)) .
\end{equation}
Note that if $X$ solves \eqref{main1} so does for all $\beta\in\R^d$
\begin{equation}
\label{transla}
X_\beta(\theta)=X(\theta+\beta)+\beta.
\end{equation}
We work mostly in Fourier space using lower case letters for the Fourier transform
\begin{equation}
x(q)=\int_{\T^d}e^{-iq\theta}X(\theta)
\non
\end{equation}
for $q\in\Z{^d}$. 
Eq. \eqref{main0} can be written as fixed point equation
\begin{equation}
\label{main}
x(q)= (\omega q)^{-2} u(q,x)\ \ q\neq 0,\ \ \ u(0,x)=0
\end{equation}
which is nontrivial since the multiplier $(\omega q)^{-2}$ is unbounded because the denominator can get arbitrary small.

For the RG we cut out small denominators $ \omega q$
in  \eqref{main} and study how the problem changes when the
cutoff is changed.  Let  $\chi:\R^+\to [0,1]$ be a smooth monotone function with
$\chi=0$ on $[0,1]$ and $\chi=1$ on $[2,\infty)$ and let
$$
\eta(t) = te^{\alpha t}
$$ 
where  $\alpha>0$ is chosen later. Define
\begin{equation}
\label{Gt }
\gamma_t(q)=\chi(\eta(t) |\omega q|)(\omega q)^{-2}.
\end{equation}
Thus $\gamma_t$ has the small denominators $|\omega q|\leq \eta(t)^{-1}$ cut off. 

We will look for a solution to the cutoff problem
\begin{equation}
\label{xt}
x(t)= \gamma_t u(x(t))
\end{equation}
by studying a bit more general one. Suppose we
can find a functional $f(t,y)$ defined in a suitable
space of maps $y:\T^d\to\R^d$ and satisfying
\begin{equation}
\label{ft}
f(t,y) = y + \gamma_t u(f(t,y)).
\end{equation}
Then $x(t)=f(t,0)$. Define also
\begin{equation}
\label{wt}
w(t,y) = u(f(t,y)).
\end{equation}
Then, differentiating \eqref{ft} in $t$ and $y$ we have
$(1-\gamma Du(f))\dot f=\dot\gamma w$ and 
$(1-\gamma Du(f))Df=1$ i.e. we should look for $f$ as a solution to
the differential equation
\begin{equation}
\label{Feq}
\dot f(t,y) = Df(t,y)\dot\gamma_t w(t,y) \, , \quad f(0,y) = y
\end{equation}
where the initial condition is due to $\gamma_0=0$.
 For $w$ we get from \eqref{wt}
 $\dot w=Du(f)\dot f=Du(f)Df\dot\gamma \ w$
 i.e. we look for the solution to
\begin{equation}
\label{EqW}
\dot w(t,y) = Dw(t,y)\dot\gamma_t w(t,y) \, , \quad w(0, y) = u(y).
\end{equation}
Our strategy then is to solve the RG equations \eqref{Feq} and
 \eqref{EqW} in a suitable space and construct the solution of
 \eqref{main} as the limit
 \begin{equation}
\label{Final}
x = \lim_{t\rightarrow \infty} f(t,0).
\end{equation}
 We need the usual Diophantine property:
for some $a,\nu>0$ and all $q\neq 0$
\begin{equation}
\label{dio}
|\omega q|\geq a|q|^{-\nu}.
\end{equation}
Then we prove
\begin{theorem}Let $v$ be real analytic. Then the equation \eqref{main0}
has a unique, up to translations \eqref{transla},  real analytic solution analytic in $\lambda$
in some neighbourhood of the origin.
\end{theorem}

\section{Identities}

The flow possesses two symmetries, exhibited in \cite{bgk},  that are crucial for the
analysis. Before defining them let us write \eqref{EqW} in the Fourier variables
\begin{equation}
\label{EqW1}
\dot w(t,q,y) = \sum_{r\in\Z{^d}}Dw(t,q,r,y)\dot\gamma_t (\omega r)w(t,r,y) 
\end{equation}
where $Dw(t,q,r,y)$ is a $d\times d$ matrix and
 setting $\kappa=\omega r$
\begin{equation}
\label{gamma}
\dot\gamma_t = \kappa^{-2} \frac{d}{dt} \chi(\eta(t) |\kappa|)
\end{equation}
$w$ and $f$ will be given by their Taylor series \begin{equation}
\label{w}
w(t,q,y)= \sum_{n=0}^{\infty} \sum_{\p \in \Z^d \times \cdots \times \Z^d} w_n(t,q, p_1, \ldots , p_n) (y(p_1), \ldots , y(p_n)).
\end{equation}  
The kernels $w_n$ are linear maps $w_n\in{\mathcal L}((\R^d)^{n}, \R^d)$,
symmetric in $p_1, \ldots p_n$; for $f$ we have a similar expansion. The equation \eqref{EqW1} becomes then a set of equations for the kernels:
\begin{equation}
\label{kernels}
\dot w_n(t, q, \p) = \sum_{k=1}^{n+1} \sum_{r\in \Z^d} k w_k(t,q,r,\p')w_{n+1-k}(t,r, \p'') \dot\gamma_t(\omega r)   \equiv
v_n(t, q, \p) 
\end{equation}
with $\p = (\p' , \p'')$. 
Eq.  \eqref{kernels} should be symmetrized in the  $\p$
but we will keep this implicit in the notation.

\subsection{ $\Z^d$ action}
To define this note first
that
at $t=0$ we get from \eqref{main1} after some calculation
\begin{equation}
\label{un}
u_n(q, \p) ={\lambda v(r)\over n!} (ir)^{\otimes^{ n+1}},
\quad
r=q-\sum p_i
\end{equation}
where from now on we identify ${\mathcal L}((\R^{d})^n, \R^d)$
with $(\R^d)^{ n+1}$.
Hence, trivially, 
\begin{equation}
\label{shift0}
u_n(q + p, p_1+p,\p') =  u_n(q , p_1,\p')
\end{equation}
for $p \in \Z^d$.
This symmetry will, for $t > 0$ take a peculiar form. Let
$$
\pi(t,q,q',y)=Dw(t,q,q',y).
$$
 Then  \eqref{shift0} can be paraphrased
as  $\tau_p\pi(0,q,q',y)\equiv \pi(0,q+p,q'+p,y)=\pi(0,q,q',y)$. On the other hand, \eqref{EqW1}
yields
\begin{equation}
\nonumber
\dot \pi(t,q,q',y) = \sum_{r}(D\pi(t,q,q',r,y)w(t,r,y) +\pi(t,q,r,y) \pi(t,r,q',y) )\dot\gamma_t (\omega r).
\end{equation}
The translated function satisfies upon shifting $r$ by $p$ in the second term and denoting $\omega p=\kappa$
\qq
\label{pi}
\tau_p\dot \pi(t,q,q',y)& =& \sum_{r}D\tau_p\pi(t,q,q',r,y)w(t,r,y)\dot\gamma_t (\omega r) \non\\
&+&\sum_{r}\tau_p\pi(t,q,r,y) \tau_p\pi(t,r,q',y)\dot\gamma_t (\omega r+\kappa) .
\qqq
Suppose now we can find a one parameter family of functions 
$\pi(t,q,q',y,\kappa)$, $\kappa\in\R$ solving eq. \eqref{pi} with
$\tau_p \pi$ replaced by $\pi(\cdot,\kappa)$.
with initial condition $\pi|_{t=0}=Du$ independently on $\kappa$. Uniqueness of solutions will then yield
\begin{equation}
\label{tra}
Dw(t,q+p,q'+p,y)=\pi(t,q,q',y,\omega p)
\end{equation}
which will be used in the second symmetry. 
The equation  \eqref{pi}  for  $\pi(\cdot,\kappa)$ can be written in terms of kernels $w_n(q,q',\p,\kappa)$,
$\p\in(\R^d)^{(n-1)}$, symmetric $\p$,
which at $n=0$ or $\kappa=0$ coincide with $w_n(q,q',\p)$. For $n>0$  eq. \eqref{kernels}
is modified to 
\begin{equation}
\label{kernels1}
\dot w_n(t,q,q', \p,\kappa)=v_n(t,q,q', \p,\kappa)
\end{equation}
with
\qq
\label{vn}
v_n(t,q,q', \p,\kappa)&=& \sum_{k+l=n+1} {\frac{k}{n}} \sum_{r} 
[(k-1) w_k(t,q,q',r,\p',\kappa)w_{l}(t,r, \p'')) \dot\gamma_t(\omega r)  \non\\
&&+  l w_k(t,q,r,\p',\kappa)w_{l}(t,r, q',\p'',\kappa)) \dot\gamma_t(\omega r+\kappa)] 
\qqq
where again symmetrization in $\p$ is implicit.   Eq. \eqref{kernels1} obviously coincides with eq.  \eqref{kernels} at $\kappa=0$, the two positions of $q'$ giving rise to symmetrization
in $(q',\p)$.
 
The $\Z^d$ action   is then given by
\begin{equation}
\label{tr1}
w_n(t,q+r, q'+r,\p, \kappa) = w_n(t,q,q', \p,\kappa + \omega r)
\end{equation}
and given $w$ satisfying \eqref{tr1} the vector field $v$ 
in  \eqref{vn} satisfies it too.
Eq \eqref{tr1} finds its use in
\begin{equation}
\label{use}
v_1(t,q,q, 0) = v_1(t,0,0, \omega \cdot q) .
\end{equation}

\subsection{ $T^d$-action}
 
Consider now $\beta \in \T^d$ and define the action $W \mapsto \tau_{\beta}W$ where
\begin{equation}
\label{action}
\tau_{\beta}W (Y)(\theta) = W(Y_{\beta})(\theta - \beta) \quad \text{with} \quad Y_{\beta}(\theta)=  Y(\theta + \beta) + \beta 
\end{equation}
From  \eqref{main} we have for $t=0$
$
\tau_{\beta} U = U$.
Moreover the vector field on the RHS of \eqref{EqW}
is invariant under $\tau_\beta$. Indeed, we have 
$$(D\tau_\beta W)(\theta,\theta',Y)=DW(\theta-\beta
,\theta'-\beta,Y_\beta)
$$ 
so the invariance follows from 
 the fact that the kernel of the operator $\dot\gamma$  in
 $\theta$ space $\dot\Gamma(\theta-\theta')$
 commutes with translations. Thus we will look for solutions
 satisfying $
\tau_{\beta} w = w$. Infinitesimally $\partial_\beta \tau_{\beta} w=0$
gives the so called \emph{Ward identity}:
\begin{equation}
\label{Ward0}
a\cdot\partial_{\theta} W(\theta,Y)= \int_{\T^d} 
DW(\theta,\theta',Y)(a+a\cdot\partial_{\theta} Y(\theta'))
 d\theta'
\end{equation}
for any $a\in\R^d$. In Fourier space this reads
\begin{equation}
\label{Ward00}
ia\cdot qw(q,y)= \sum_{q'}
Dw(q,q',y)(a\delta_{q'0}+ia\cdot q' y(q'))
\end{equation}
or, in terms of the kernels
\begin{equation}
\label{Wardn}
ia\cdot qw_n(q,\p)= (n+1)w_{n+1}(q,0,\p)a+i\sum_i a\cdot p_i
w_n(q,\p) 
\end{equation}
The $n=0$ relation will play a crucial role later:
\begin{equation}
\label{Ward2}
i w_0(q) \otimes q= w_1(q,0).
\end{equation}
Furthermore, using Eq. \eqref{EqW} together with \eqref{Ward2} we get
\begin{equation}
\label{dotWard}
\dot w_0(0) = \sum_{r} w_1(0,r)\dot\gamma(r)w_0(r) = -i \sum_r r \otimes w_0(-r)\dot\gamma(r)w_0(r) = 0
\end{equation}
since the summand is antisymmetric in $r$. Thus, since $w_0(0)|_{t=0} = u_0(0) = \int \lambda \partial_{\theta} v(\theta) d\theta = 0$, Eq. \eqref{dotWard} implies 
\begin{equation}
\label{Ward1}
w_0(0) = 0.
\end{equation}
To see the consequences of eqs. \eqref{Ward1} and  \eqref{Ward2} note that eq. \eqref{un} implies
\begin{equation}
\label{note}
u_1(q,q',y) = u_1(-q',-q,y)^T.
\end{equation}
It is readily checked $\tilde w_m(q,q',\p,\kappa)\equiv w_m(-q',-q,\p,-\kappa)^T$
satisfies equation \eqref{kernels1} as well.
Then, given uniqueness of solutions to  eq. \eqref{kernels1} we conclude
\begin{equation}
\label{note1}
w_n(q,q',\p,\kappa) = w_n(-q',-q,\p,-\kappa)^T.
\end{equation}
Consider in particular eq.   \eqref{kernels1}  for $n=1$  and $q=q'=0$.
By  \eqref{note1}  $w_2(t,0,0,\p,\kappa)$ is even in $\kappa$. Hence
\begin{equation}
\label{fin}
\dot w_1(0,0,\kappa)=
\sum_{r} w_1(0,r,\kappa)w_{1}(0, -r,-\kappa))^T \dot\gamma_t(\omega r+\kappa)+{\rm even}
\end{equation}
where  \eqref{note1} was used. Since $\dot\gamma_t$ is even, we deduce that the RHS is even in $\kappa$ and therefore $w_1(0,0,\kappa)$ is also even.
We will show below that $w_n$ are $C^2$ in $\kappa$. Hence
\begin{equation}
\label{w3}
\partial_\kappa w_1(0,0,0)=0.
\end{equation}

\section{Estimates}
We describe now the functional space where \eqref{kernels1} is solved. Consider first $u$ given by \eqref{un}.
By assumption, $v$ is analytic in   $|\Im\theta_i|\leq 2b$
for some $ b>0$. Then eq.
 \eqref{un} implies
 \begin{equation}
\label{u}
|u_n| \leq  C|\lambda|R^{n}e^{-b|r|}
\end{equation}
for some $R$ depending on $b$. 
Given $\q \in (\Z^{d})^n$ and $\kappa \in \R$, let $w_n(\q, \kappa)\in (\R^d)^{\otimes^{n+1}}$ be symmetric under simultaneous exchanges of $q_i$ and $q_j$ and the corresponding $\R^d$'s for $i,j\geq 3$,
for $\kappa=0$ for $i,j\geq 2$ and satisfying \eqref{Wardn},\eqref{Ward1} and \eqref{note1}. Let
\begin{equation} 
\label{Lambda}
\Lambda_t = \{ q \in \Z^d \, | \, \eta(t) |\omega \cdot q| \leq 3 \}
\end{equation} 
and, for $\beta>0$,
\begin{equation}
\label{beta}
\beta_t=\hf(1+ \frac{1}{1+t})\beta
\end{equation}
Define
\begin{equation}
\label{normt}
\Vert w_n \Vert_t = \sup_{\p \in \Lambda_t^n} \sum_{q \in \Lambda_t} e^{\beta_t |q- \sum p_i|}|w_n(q,\p, \cdot)|_t
\end{equation}
where $|\cdot|_t$ is the following $\mathcal{C}^2$ norm: 
\begin{equation}
\label{normC2}
|f|_t = \sup_{\eta(t)|\kappa| \leq 1} \sum_{i=0}^2 \eta(t)^{-i}|\partial_i f(\kappa)|
\end{equation}
and $|\cdot|$ is the standard norm in $(\R^d)^{\otimes^{n+1}}$.

Let $w(t)= \{w_n(t)\}_{n \in \N}$ and define
\begin{equation}
\label{norm}
\Vert w \Vert = \sup_{t \geq 0} \left( e^{2t} \Vert w_0(t) \Vert_t + \eta(t)^2\Vert w_1(t) \Vert_t + \sup_{n > 1}n^2 \rho^ne^{\left(\frac{3}{2}-n\right)t}\Vert w_n(t) \Vert_t \right).
\end{equation}
We have then 
\begin{proposition} 
\label{mainprop} 
The equation \eqref{EqW} has a unique solution 
analytic in $\lambda$ in some ball around zero and satisfying
\begin{equation}
\label{bound0}
\Vert w \Vert \leq C(v) |\lambda|
\end{equation}
where the constant $C$ depends on $v$. 
\end{proposition}
\begin{proof}
We write \eqref{kernels1} as a fixed point equation
\begin{align}
\label{integral}
w(t)=u+\int_0^t v(s)ds\equiv u+\Phi(t,w)
\end{align}
where $v_n$ is given by  \eqref{kernels} for $n=0$
and by \eqref{vn} for $n>0$.

We start by showing
\begin{equation}
\label{unorm}
\Vert u \Vert\leq C\lambda.
\end{equation}
We use the bound \eqref{u} and take $\beta<b$ 
to get
\begin{equation}
\label{unormn}
\Vert u_n \Vert_0\leq CR^{n}|\lambda|.
\end{equation}
 For $n=0$ we have
\begin{equation}
\label{u0}
\sum_{q \in \Lambda_t} |u_0(q)|e^{\beta_t |q|}  \leq \sup_{0 \neq q \in \Lambda_t}e^{-(\beta - \beta_t)|q|} \Vert u_0 \Vert_0
\end{equation}
where $q \neq 0$ since by  \eqref{un} $u_0(0)= 0$. 
By the Diophantine condition eq. \eqref{dio}  $|\omega \cdot q| \eta(t) < 3$ implies $|q| \geq \left( \frac{a}{3}\eta(t) \right)^{\frac{1}{\nu}}$. Since $\beta - \beta_t = \frac{1}{2} \beta \frac{t}{1+t}$ the sup is superexponential in $t$, hence
\begin{equation}
\label{u0ok}
\Vert u_0 \Vert_t \leq C |\lambda|e^{-2t}.
\end{equation}
For $n=1$, we get from \eqref{Ward2} $u_1(q,q) = 0$, so
\begin{equation}
\label{u1}
\sum_{q \in \Lambda_t} |u_1(q,p)|e^{\beta_t |q-p|} \leq \sup_{\substack{q,p \in \Lambda_t\\q \neq p}} e^{-(\beta - \beta_t)|q-p|}\Vert u_1 \Vert_0
\end{equation}
and since $\eta(t) |\omega \cdot (q-p)| \leq 6$ 
we get as in \eqref{u0ok}
\begin{equation}
\label{u1ok}
\Vert u_1\Vert_t \leq CR |\lambda| \eta(t)^{-2}.
\end{equation}
Finally for $u_n$, $n > 1$ we just use   $\beta_t \leq \beta_0$ to conclude
\begin{equation}
\label{unok}
\sum_{q \in \Z^d} |u_n (q,\p)| e^{\beta_t |q-\p|} \leq \Vert u_n\Vert_0
\leq CR^ne^{(n- \frac{3}{2})t} |\lambda|\leq Cn^{-2}\rho^{-n}e^{(n- \frac{3}{2})t} |\lambda| 
\end{equation}
provided we take $\rho R<1$. Then Eq.  \eqref{unorm} follows from
 \eqref{u0ok}, \eqref{u1ok}, \eqref{u1ok} and \eqref{norm}.

Thus it suffices to prove that $\Phi$ is a contraction from the ball $B_u(R|\lambda|)$ to
$B_0(R|\lambda|)$ for a suitable $R$. 
We shall treat separately $\Phi_n$ for $n=0, n=1$ and $n \geq 2$.

\noindent $n=0$.  From \eqref{kernels} we have
\begin{equation}
\label{v0}
v_0(s,q)=\sum_{r \in \Lambda_s} w_1(s,q,r)
w_0(s,r)\dot\gamma_s(\omega r)
\end{equation}
where we may restrict $r\in\Lambda_s$ since by \eqref{gamma}
 $\dot\gamma_s(x)$ is supported in $\eta(s)|x| \in [1,2]$.
 Using $e^{\beta_t |q|} \leq e^{\beta_t|q-r| + \beta_t|r|}$ and  $\beta_t\leq\beta_s$
together with the obvious 
\begin{equation}
\label{obv}
|\cdot |_t \leq |\cdot |_s, \ \ |AB|_s \leq c|A|_s|B|_s
\end{equation}
we get
\begin{equation}
\sum_{q \in \Lambda_t} \!\!e^{\beta_t |q|}|\Phi_0(t,q)|_t 
\leq C \sum_{q \in \Lambda_t} \!  \int_0^t\sum_{r \in \Lambda_s}
 \!\!e^{\beta_s|q-r|}|w_1(s,q,r)|_s e^{\beta_s|r|}|w_0(s,r)|_s |\dot\gamma_s|_s e^{-(\beta_s - \beta_t)|q|}ds.
\label{phi0}
\end{equation}
From \eqref{gamma} we infer
\begin{equation}
\label{gammabound}
|\dot\gamma_s|_s \leq C \eta(s)^2
\end{equation}
so
\begin{equation}
\label{so}
\Vert \Phi_0(t)\Vert_t \leq C \Vert w \Vert^2 \int_0^t e^{-2s} \inf_{0 \neq q \in \Lambda_t} e^{-(\beta_s-\beta_t)|q|}ds.
\end{equation}
Since $\beta_s-\beta_t\geq\hf\beta(1+t)^{-2}(t-s)$ the Diophantine
property implies
\begin{equation}
\label{dp}
e^{-(\beta_s - \beta_t)|q|} \leq Ce^{-(t-s)a(t)}
\end{equation}
where $a(t) \rightarrow \infty$ as $t \rightarrow \infty$. Hence 
\begin{equation}
\label{phi0ok}
\Vert \Phi_0(t)\Vert_t \leq C e^{-2t} \Vert w \Vert^2.
\end{equation}

\noindent $n=1$. 
Write $v_1=v_a+v_b$ with
\qq
\label{va}
v_a(s,q,q',\kappa)  &=&   \sum_{r\in \Lambda_s}  w_1
(s,q,r) w_1(s,r,q') \dot\gamma_s(\omega r + \kappa)\\
v_b(s,q,q',\kappa)  &=&   \sum_{r\in \Lambda_s} 
w_2(s,q,r,q',\kappa) w_0(s,r) \dot\gamma_s(\omega r) \label{vb}
\qqq
and $\Phi_1$ similarily. We could restrict $r\in \Lambda_s$
also in \eqref{va} since 
 $\eta(t)|\kappa| \leq 1$ and so $\eta(s) |\omega  r| \leq 2 + \eta(s)|\kappa| \leq 3  $
because $\eta(t)$ is increasing in $t$.

We consider first the case  $q \neq q'$
and proceed as in the $n=0$ case. 
For
 $v_a$ we insert
\begin{equation}
\non
e^{\beta_t|q-q'|} \leq e^{\beta_s \left[ |q-r| + |r-q'|  \right]} e^{-(\beta_s - \beta_t)|q-q'|}
\end{equation}
and use \eqref{dp}, \eqref{norm},\eqref{obv} and
 \eqref{gammabound} to get 
\begin{equation}
\label{insert2}
\sum_{q \neq q'} |e^{\beta_t|q-q'|} \Phi_{a}(t,q,q')|_t \leq C \Vert w \Vert^2 \int_0^t \eta(s)^{-2}
 e^{-(t-s)a(t)}ds \leq C \eta(t)^{-2} \Vert w \Vert^2
\end{equation}
Similarily, for $\Phi_{b}$ we use
\begin{equation}
\non
e^{\beta_t |q-q'|} \leq C e^{\beta_s (|q-r-q'| + |r|)} 
e^{-(\beta_s - \beta_t)|q-q'|}
\end{equation}
to get
\begin{align}
\label{phi12}
\sum_{q \neq q'}e^{\beta_t |q-q'|}  |\Phi_{b}(t,q,q')|_t  \leq C \Vert w \Vert^2  \int_0^t e^{s/2 - 2s} \eta(s)^2 e^{-(t-s)a(t)}
\leq C \eta(t)^{-2} \Vert w \Vert^2
\end{align}
provided we take $2\alpha\leq {3\over 2}$ in the definition
of $\eta(t)$.

Let now $q=q'$. Call the summands in \eqref{va} and
\eqref{vb} $v_{ar}$ and $v_{br}$. We consider first in the integral
in \eqref{integral} $s\geq s_0$ with   
$ \eta(s_0)^{-1} = 4\eta(t)^{-1}$. 
Using
\begin{equation}
\label{var}
|v_{ar}(s,q,q)|_t\leq Ce^{-2\beta_s|q-r|}\eta(s)^{-2}\Vert w \Vert^2
\end{equation}
we get
\begin{equation}
\label{vat}
\int_{s_0}^t |v_a(s,q,q) |_tds\leq C \eta(t)^{-2}\Vert w \Vert^2.
\end{equation}
In the same way we get
\begin{equation}
\label{vbt}
\int_{s_0}^t |v_{b}(s,q,q)|_t\leq C \Vert w \Vert^2 \int_{s_0}^t e^{s/2-2s}\eta(s)^{2}
ds \leq C \eta(t)^{-2} \Vert w \Vert^2
\end{equation}

Finally, consider $v_a(s,q,q)$ for $s\leq s_0$. For such $s$
the $r=q$ term is zero. Indeed,  
$\supp \dot\gamma_s \subset [\eta(s)^{-1}, 2\eta(s)^{-1}]$, but 
$|\omega  q + \kappa| \leq 4\eta(t)^{-1}\leq\eta(s)^{-1}$. 
For $r\neq q$ we have 
\begin{equation}
\label{dg}
|\partial^i_{\kappa} v_{ar}| \leq Ce^{-2\beta_s|q-r|}\eta(s)^{-2+i}\Vert w \Vert^2
\end{equation}
for $i \leq 2$ and $|\kappa| \leq \eta(s)^{-1}$. 
We conclude, since $|q-r|>C\eta(s)^{1\over \nu}$ that
\begin{equation}
\label{iva}
| \partial^i_\kappa \int_0^{s_0} v_a(s,q,q)ds | \leq C \Vert w \Vert^2 \end{equation}
which holds for $|\kappa| \leq 4\eta(t)^{-1}$ since
$ \eta(s_0)^{-1}\geq 4\eta(t)^{-1}$. 
In a similar manner we bound, for $|\kappa| \leq 4 \eta(t)^{-1}$,
\begin{equation}
\label{ivb}
| \partial^i_\kappa \int_0^{s_0} v_b(s,q,q)ds | 
\leq C \Vert w \Vert^2
\end{equation}
By the symmetry \eqref{use}
\begin{equation}
\label{symm}
v_1(s,q,q, \kappa) = v_1(s,0,0; \omega q + \kappa).
\end{equation}
Let $\Psi(q,\kappa)=\int_0^{s_0}v_1(s,q,q,\kappa)$. 
Since $\partial_\kappa^i\Psi(0,0)=0$ for $i<2$ we have
\begin{equation}
\label{Dint}
\Psi(q,\kappa)=\int_0^{\omega \cdot q + \kappa} 
\partial^2_{\kappa}\Psi (0, \kappa')(\omega q+\kappa-\kappa' )d\kappa'.
\end{equation}
Since $\Psi(q,\kappa)$ 
satisfies \eqref{iva} for $|\kappa| \leq 4 \eta(t)^{-1}$
and  $|\omega \cdot q + \kappa| \leq 3 \eta(t)^{-1}+ \eta(t)^{-1}$, \eqref{Dint} is defined and by \eqref{iva} we infer that
\begin{equation}
\label{Dbound}
|\Psi|_t \leq C \Vert w \Vert^2 \eta(t)^{-2}.
\end{equation}
Thus  \eqref{insert2},  \eqref{phi12} and  \eqref{Dbound} give   
\begin{equation}
\label{phi1final}
\Vert \Phi_1\Vert_t \leq C \eta(t)^{-2} \Vert w \Vert^2.
\end{equation}

To finish the proof we need to deal with $n >1$ in \eqref{integral}. Very crude bounds suffice. Using
\begin{equation}
\label{exp}
e^{\beta_t(q-q' \sum p_i)} \leq e^{\beta_s(|q-r-q' - \sum p'_i | + |r - \sum p''_i|)}
\end{equation}
for the first term and $|q-r-\sum p'_i | + |r -q'- \sum p''_i|$
in the exponent in the second term
we get
\begin{align}
\label{phin}
\sum_q|\Phi_n|_t e^{\beta_t |q-q'- \sum p_i|} & \leq \sum_{k=1}^{n+1} k \int_0^t \Vert w_k\Vert_s \Vert w_{n+1-k}\Vert_s \eta(s)^2 ds
 \leq C \Vert w \Vert^2\rho^{-n} \int_0^t   [ \eta(s)^{-2} n^{-2} e^{(n - 3/2)s}  \notag \\
 &+ (n+1)^{-1}\rho^{-1}  e^{(n - 1/2)s}e^{-2s} \quad+\rho^{-1} \sum_{k=2}^n \frac{1}{k} \frac{1}{(n+1-k)^2} e^{(n-2)s}] \eta(s)^2 ds \notag \\
& \leq C \Vert w \Vert^2 \rho^{-n-1}e^{(n-3/2)t} n^{-2}[ \frac{\rho}{n} + 
1 + n\sum_{k=2}^n \frac{1}{k} \frac{1}{(n+1-k)^2} ]
\end{align}
provided $\eta(s)^2 e^{-s/2} < C$ which holds if 
$\alpha < \frac{1}{4}$. The term in the squared parenthesis is bounded in $n$ and thus
\begin{equation}
\label{phinfinal}
\Vert \Phi_n \Vert_t \leq C n^{-2}\rho^{-n-1}e^{(n-3/2)t} \Vert w \Vert ^2.
\end{equation}
Eqs. \eqref{phi0ok}, \eqref{phi1final}, \eqref{phinfinal} imply 
 that $\Phi$ maps the ball $B_u( r)$ to  $B_0( r)$ 
 with $r=
 C\rho^{-1}\lambda^2$ if
   $\rho^{-1}\lambda$ is small enough. Contraction
 in this ball is similar.

\end{proof}

\section{The solution}

Let us finally solve the $f$-equation \eqref{Feq}, written as
\begin{equation}
\label{Eqf}
f_n(t,q, \p)  = a\delta_{n,1} \delta_{q,p} \mathbbm{1} + \sum_{k=1}^{n+1} \int_0^t \sum_{q'} k f_k(s,q,q',\p') \dot\gamma_s(\omega \cdot q') w_{n+1-k}(s,q',\p'') ds
 \equiv g_n +  \F_n
\end{equation}
where $a=1$ for  \eqref{Feq} but we keep $a\geq 0$
for later purposes. Let
\begin{equation}
\label{fnormt}
\Vert f_n \Vert_t = \sup_{\p \in \Lambda_t^n} \sum_{q \in \Z^d} e^{\beta_t |q- \sum p_i |}|f_n(q, \p)|
\end{equation}
and
\begin{equation}
\label{fnorm}
\Vert f \Vert = \sup_{t \geq 0} \left( \Vert f_0 \Vert_t + \Vert f_1\Vert_t + \sup_{n \geq 2} n^{2} \rho^{n}e^{(1-n)t} \Vert f_n \Vert_t \right).
\end{equation}
Note that \eqref{fnormt} differs from \eqref{normt} in the 
constraint on $q$ and the absence of $\kappa$. We use abusively
the same notation. We have
\begin{proposition} \label{final}
Equation \eqref{Eqf} has a unique solution in the ball $\Vert f - g \Vert \leq Ca \rho^{-1}\Vert w \Vert$.
\end{proposition}

\begin{proof}
We proceed as in the proof of Proposition \ref{mainprop} to get
\begin{equation}
\label{Fhigh}
\Vert \F_{n} \Vert_t \leq \sum_{k=1}^{n+1} \int_0^t k \Vert f_{k}(s) \Vert_s \Vert \dot\gamma_s \Vert \Vert w_{n+1-k} \Vert_s ds.
\end{equation}
For $n=0$ this gives
\begin{equation}
\label{Fzero}
\Vert \F_{0}\Vert \leq C \Vert w \Vert \Vert f \Vert \int_0^t \eta(s)^2 e^{-2s} ds \leq C \Vert w \Vert \Vert f \Vert
\end{equation}
and for $n > 1$
\begin{align}
\Vert \F_{n} \Vert_t & \leq  C\rho^{-n-1} \Vert w \Vert \Vert f \Vert
\int_0^t  \Big[\rho n^{-2}e^{(n-3/2)s} + (n+1)^{-1}e^{ns}e^{-2s} 
 + \rho n^{-1}e^{(n-1)s}\eta(s)^{-2} \notag \\
& \quad+ \sum_{k=2}^{n-1} k^{-1} e^{(k-1)s} e^{(n+1-k-3/2)s}\Big]\eta(s)^2ds 
 \leq  C\Vert w \Vert \Vert f \Vert \rho^{-n-1}n^{-2}e^{(n-1)t}. \label{Fmorethanone}
\end{align}
For $n=1$, the $k=2$ term in \eqref{Fhigh} is bounded by
\begin{equation}
\label{Fzero}
 C\rho^{-1} \Vert w \Vert \Vert f \Vert \int_0^t  e^{s}e^{-2s} \eta(s)^2ds \leq C\rho^{-1}  \Vert w \Vert \Vert f \Vert.
\end{equation}
For the $k=1$ term we need to improve from \eqref{Fhigh}
a bit and extract as before the factor $e^{(\beta_t - \beta_s)|q-q'|} $,
bounded by 
$ e^{-(t-s)\alpha(s)}$ to get the bound
\begin{equation}
\label{Fone}
C \Vert w \Vert \Vert f \Vert \int_0^t  \eta(s)^{-2} \eta(s)^2e^{-(t-s)\alpha(s)}ds \leq C\Vert w \Vert \Vert f \Vert
\end{equation}
for the $q=p\neq q'$ terms in the eq. \eqref{Eqf}. For  $q=p=q'$ 
we observe that $\dot\gamma_s(\omega q)$ vanishes for
$3\eta(s)\leq\eta(t)$ since $q\in \Lambda_t$. Thus the integral
in \eqref{Eqf} is restriceted to $s>t-{\mathcal O}(1)$ and therefore
bounded by  \eqref{Fone} again.

So, altogether we have
\begin{equation}
\label{F1}
\Vert \F_{1} \Vert_t \leq C\rho^{-1}  \Vert w \Vert \Vert f \Vert.
\end{equation}

\end{proof}
We may now prove Theorem 1. First we note the limit \eqref{Final}
exists and is real analytic. Indeed, let $x(t,q)=f_0(t,q)$. From Proposition 2 we deduce
\begin{equation}
\label{conv}
\Vert f_0(t)-f_0(s)\Vert_t ={\mathcal O}(e^{-2s}\eta(s)^2)
\end{equation}
i.e. $x(t,q)$ converge uniformly to an exponentially decaying
sequence $x(q)$.  We will shortly prove  that $x(t)$ solves the equations
\begin{equation}
\label{apr}
x(t)=\gamma_tu(x(t)), \ \ u(x(t))=w(t,0).
\end{equation}
The first equation implies that $x(t,0)=0$ and for $|\omega q|\geq 2\eta(t)^{-1}$, 
$(\omega q)^2x(t,q)=u(q,x(t))$. Hence, for $q\neq 0$
\begin{equation}
\label{apr1}
(\omega q)^2x(q)=u(q,x)
\end{equation}
and $x(0)=0$.
The second equation \eqref{apr} combined with  \eqref{Ward1}
yields $u(0,x(t))=0$ and so by limits the second eq. \eqref{main} follows. Our solution is clearly analytic in $\lambda$ and
thus unique (up to the translations) since the equation
\eqref{main0} determines uniquely the Taylor coefficients
of its analytical solution.

We still need to prove \eqref{apr}. Consider the functionals
\begin{equation}
\label{phi}
\phi(t,y)=y+\gamma_tu(f(t,y)),\ \ \ \psi(t,y)=u(f(t,y)).
\end{equation}
$\phi$ and $\psi$ are analytic in $y$ in a neighbourhood of $0$
in $\ell^1(\Lambda_t)$. 
It suffices to show $\phi=f$ and $\psi=w$ since  \eqref{apr}
follows at $y=0$. Differentiating and using  \eqref{Feq}
we get
\begin{equation}
\label{phidot}
\dot\phi=\dot\gamma\psi+\gamma Du\dot f=\dot\gamma\psi+
\gamma DuDf\dot\gamma w=D\phi\dot\gamma w+\dot\gamma\rho
\end{equation}
where $\rho=\psi-w$.
Similar calculation gives $ \dot\psi=D\psi\dot\gamma w$
and combining with  \eqref{EqW} we get
\begin{equation}
\label{rhodot}
\dot\rho=D\rho\dot\gamma w.
\end{equation}
Initial conditions are $\phi(0,y)=f(0,y)$ and $\rho(0,y)=0$.
Applying Proposition 2 with $a=0$ we get that $\rho=0$
and then with  $a=1$  that $\phi=0$. Hence
\eqref{apr} follows.

\end{document}